\documentclass[12pt,tightenlines,eqsecnum,floats,aps,amsmath,amssymb,nofootinbib,prd,showpacs]{revtex4}

\usepackage{setspace}
\usepackage{amsmath,amssymb,amsfonts,amsthm}
\usepackage{graphicx}
\usepackage{enumerate} 
\usepackage{colordvi} 

\def\be{\begin{equation}}
\def\ee{\end{equation}}
\def\ba{\begin{eqnarray}}
\def\ea{\end{eqnarray}}
\def\bi{\begin{itemize}}
\def\ei{\end{itemize}}
\def\bnum{\begin{enumerate}}
\def\enum{\end{enumerate}}

\def\V{\mathcal{V}}
\def\v{\nu}
\def\p{\phi}
\def\H{{\cal H}}

\def\Hk{\H_{\rm kin}}
\def\Hp{\H_{\rm phy}}
\def\h{\hat }
\def\b{\bar}
\def\ub{\underbar}
\def\ul{\underline}

\def\f{\frac}

\def\lp{{\ell}_{\rm Pl}}
\def\l{\lambda}
\def\lo{\ell_o}
\def\bra{\langle}
\def\ket{\rangle}

\def\R{\mathbb{R}}

\def\I{\mathcal{I}}

\def\dd{\textrm{d}}

\def\utw#1{\rlap{\lower1ex\hbox{$\sim$}}#1{}} 

\usepackage{enumerate}

\usepackage{colordvi}
\newcounter{mnotecount}[section]

\newcommand{\comment}[1]{}
\begin{document}

\title{Loop Quantum Cosmology and Spin Foams}

\author{Abhay Ashtekar}\email{ashtekar@gravity.psu.edu}
\author{Miguel Campiglia} \email{miguel@gravity.psu.edu}
\author{Adam Henderson}\email{adh195@psu.edu}
 \affiliation{Institute for Gravitation and the Cosmos \& Physics
 Department,
 Penn State, University Park, PA 16802-6300, U.S.A.}

\begin{abstract}

Loop quantum cosmology (LQC) is used to provide concrete evidence in
support of the general paradigm underlying spin foam models (SFMs).
Specifically, it is shown that: i) the physical inner product in the
timeless framework equals the transition amplitude in the
deparameterized theory; ii) this quantity admits a 
vertex expansion a la SFMs in which the $M$-th term refers just to
$M$ volume transitions, without any reference to the time at which
the transition takes place; iii) the exact physical inner product is
obtained by summing over just the discrete geometries; no `continuum
limit' is involved; and, iv) the vertex expansion can be interpreted
as a perturbative expansion in the spirit of group field theory.
This sum over histories reformulation of LQC also addresses certain
other issues which are briefly summarized.

\end{abstract}


\maketitle

\section{Introduction}
\label{s1}
\bigskip

In his celebrated Ph.D. work Feynman \cite{rpf} began with the
Hilbert space formulation of non-relativistic quantum mechanics and
reformulated the unitary evolution generated by the Hamiltonian as a
sum over histories. By mimicking the spirit of that procedure,
Reisenberger and Rovelli \cite{rr} obtained a sum over histories
formulation of loop quantum gravity (LQG). Through contributions of
many researchers it has become an active research area that goes
under the name `spin foam models' (SFMs) (see, e.g., \cite{sfrev}).
In LQG, spin network states can be used to construct a convenient
orthonormal basis in the kinematical Hilbert space. A key challenge
is to extract physical states from them by imposing constraints.
Formally this can be accomplished by the group averaging procedure
which also provides the physical inner product between the resulting
states. A primary goal of SFMs is to provide a path integral to
realize this idea.

Heuristically, the main idea behind this construction can be
summarized as follows. Consider a 4-manifold $M$ bounded by two
3-surfaces, $S_1$ and $S_2$, and a simplicial decomposition thereof.
One can think of $S_1$ as an `initial' surface and $S_2$ as a
`final' surface. Fixing a spin network on each of these surfaces
specifies an `initial' and a `final' quantum geometry. A quantum
4-geometry interpolating between the two can be constructed by
considering the dual triangulation of $M$ and coloring its surfaces
with half integers $j$ and edges with suitable intertwiners. The
idea is to obtain the physical inner product between the two states
by summing first over all the colorings for a given triangulation,
and then over triangulations keeping the boundary states fixed. (The
second sum is often referred to as `vertex expansion' because the
$M$-th term in the series corresponds to a dual triangulation with
$M$ vertices.) Since each triangulation with a coloring specifies a
quantum geometry, the sum is regarded as a path integral over
physically appropriate 4-geometries. In ordinary quantum mechanics
and Minkowskian field theories where we have a fixed background
geometry, such a path integral provides the (dynamically determined)
transition amplitude for the first state, specified at initial time,
to evolve to the second state at the final time. In the background
independent context of quantum gravity, a priori one does not have
access to a time variable and dynamics is encoded in constraints.
Therefore the notion of a transition in a pre-specified time
interval fails to be meaningful unless the constraints are solved
via deparametrization, using a relational time variable. Rather, the
sum over histories now gives the physical inner product between
solutions to quantum constraints.

Over the last two years there have been significant advances in
SFMs. In particular, relation to the kinematics underlying LQG has
now been clearly established \cite{epr,fk,eprl,kkl}. However,
while the path integral is well-motivated by exploiting an
interplay between general relativity and the well-understood BF
theory, it requires a new key ingredient ---the vertex amplitude.
While this amplitude \emph{is} severely constrained by several
requirements, it has not been systematically derived following
procedures used in well-understood field theories, or, starting
from a well-understood Hamiltonian dynamics. Therefore a number of
issues still remain. For example, the final `vertex expansion' is
a discrete sum in which each term is a sum over colorings, for a
fixed triangulation. A priori it is somewhat surprising that the
answer can be written as a \emph{discrete} sum. Would one not have
to take a continuum limit at the end as in Minkowskian field
theories? The physical inner product resembles the transition
amplitude used in standard quantum field theories. Is there a
deeper meaning in this resemblance? Loop quantum cosmology (LQC)
provides a physically interesting yet technically simple context
to explore such open issues because one can arrive at a sum over
histories starting from a fully controlled Hamiltonian theory
\cite{lqcrev}. The goal of this communication is to show that a
detailed analysis of this example provides strong support for the
general paradigm that underlies SFMs and also sharpens the
discussion of some open issues. We will sketch the general ideas
and summarize results, leaving the detailed derivations and
discussions to a longer article \cite{ach}.

\section{LQC Dynamics as a Sum Over Histories}
\label{s2}

In this communication we will focus on the simplest LQC model that
has been analyzed in detail \cite{apslett, aps1,aps2,acs}: the
k=0, $\Lambda$=0 Friedmann model with a massless scalar field as a
source. However, it is rather straightforward to extend this
analysis to allow for a non-zero cosmological constant or
anisotropies or to the spatially compact k=1 case. 

In the Hamiltonian theory one begins by fixing a (spatial) manifold
$S$, topologically $\R^3$, cartesian coordinates $x^i$ thereon, and
a fiducial metric $q^o_{ab}$ given by $q^o_{ab} \dd x^a \dd x^b =
\dd x_1^2 + \dd x_2^2 + \dd x_3^2$. The physical 3-metric $q_{ab}$
is then determined by a scale factor $a$; \, $q_{ab} = a^2
q^o_{ab}$. For the Hamiltonian analysis one fixes a cubical fiducial
cell $\V$ whose volume with respect to $q_{ab}$ is given by $V = a^3
V_o$. The quantity $\v$ defined by $V = 2\pi\gamma \lp^2\, |\v|$
turns out to be a convenient configuration variable, where $\gamma$
is the Barbero-Immirzi
parameter \cite{acs}.%
\footnote{In LQG the basic geometric variable is an orthonormal
triad and the physical metric $q_{ab}$ is constructed from it. If
the triad has the same orientation as the fiducial one, given by the
coordinates $x^i$, the configuration variable $\v$ is positive and
if the orientations are opposite, $\v$ is negative. Physics of the
model is insensitive to the triad orientation and hence to the sign
of $\v$. In particular quantum states satisfy $\Psi(\v,\p) =
\Psi(-v, \p)$.}

The kinematical quantum states are functions $\Psi(\v,\p)$ with
finite norm $||\Psi||^2 := \sum_{\v}\, \int \dd\p\,
|\Psi(\v,\p)|^2$. Thus, $\Psi$ can have support only on a countable
number of points on the $\v$-axis and a (generalized) orthonormal
basis in the kinematical Hilbert space $\Hk$ is given by
$|\v,\p\ket$ with
\be \label{kin} \bra \v^\prime,\p^\prime\,|\, \v,\p\ket_{\rm kin}
= \delta_{\v^\prime,\,\v}\,\delta(\p^\prime, \p)\, .\ee
To obtain the physical Hilbert space, one first notes that the
quantum constraint can be written as $\partial^2_\p \Psi(\v,\p) = -
\Theta \Psi(\v,\p)$ where $\Theta$ is a second order difference
operator
\ba \label{theta} \big(\Theta \Psi\big)(\v) := - \f{3\pi
G}{4\lo^2}\, \Big[\!\!&&\!\! \sqrt{|\v(\v+4\lo)|}\, (\v+2\lo)\,
\Psi(\v+4\lo)\,-\, 2\v^2\Psi(\v)\nonumber\\
&+&\, \sqrt{|\v(\v-4\lo)|}\, (\v-2\lo)\, \Psi(\v-4\lo)\, \Big] \ea
where $\lo$ is related to the `area gap' $\Delta =
4\sqrt{3}\pi\gamma\, \lp^2$ via $\lo^2 = \,\Delta$. $\Theta$ turns
out to be a positive and self-adjoint operator on the
gravitational part of the kinematic Hilbert space. The form of
$\Theta$ shows that the space of solutions to the quantum
constraint can be naturally decomposed into sectors in which the
wave functions have support on specific `$\v$-lattices'
\cite{aps2}. For definiteness, we will restrict ourselves to the
lattice $\v = 4n\lo$ where $n$ is an integer. Details of the
expression of $\Theta$ will not be needed in this analysis.

The similarity of the form of the quantum constraint to the
Klein-Gordon equation suggests that we use $\p$ as relational time
to deparameterize the quantum theory. As in the Klein-Gordon
theory, one can perform a group averaging procedure to arrive at
the physical Hilbert space $\Hp$ \cite{aps1}. Elements
$\Psi(\v,\p)$ of $\Hp$ are solutions to the `positive frequency'
quantum constraint equation
\be\label{sch} -i\partial_\p\,\Psi (\v,\p) = \sqrt{\Theta}
\Psi(\v,\p)\, \equiv\, H \Psi(\v,\p)\ee
with a finite norm
\be \label{norm}||\Psi||^2_{\rm phys} = \sum_{\v=4n\lo}\,
|\Psi(\v,\p=\p_0)|^2\, .\ee
Thanks to (\ref{sch}) the norm is independent of the `time instant'
$\p_0$ at which it is evaluated. Note that, because of
deparametrization, the quantum constraint can be regarded as
`evolving the state in relational time $\p$'.

Because of the close similarity of (\ref{sch}) with the
Schr\"odinger equation, we can now pass to a sum over histories
formulation of quantum dynamics a la Feynman. The object of interest
is the transition amplitude $A(\v_f,\p_f;\, \v_i,\p_i)$ to go from
the configuration $\v_i$ at an initial time $\p_i$ to $\v_f$ at the
final time $\p_f$. It is immediate from the form of (\ref{sch}) that
this amplitude depends only on the difference $\p_f-\p_i$.
Therefore, without loss of generality we will set $\p_i=0$ and $\p_f
= \varphi$ and refer to this time interval as $\I$. Let us divide
$\I$ into $N$ parts each of length $\epsilon = \varphi/N$ and write
the transition amplitude as
\ba \label{lqc} A(\v_f,\varphi;\,\v_i,0) := \bra \v_f|
e^{iH\varphi} |\v_i\ket &=& \sum_{\b\v_{N-1},\ldots ,\b\v_{1}}\,\,
\bra\b\v_N|e^{i\epsilon H}|\b\v_{N-1}\ket
\bra\b\v_{N-1}|e^{i\epsilon H}|\b\v_{N-2}\ket\,\dots\,
\bra\b\v_{1}|e^{i\epsilon H}|\b\v_{0}\ket \nonumber\\ &=&
\sum_{\b\v_{N-1},\ldots ,\b\v_{1}}\,\, U_{\b\v_N\b\v_{N-1}}\,
U_{\b\v_{N-1}\b\v_{N-2}}\, \ldots \,U_{\b\v_1\b\v_0} \, ,\ea
where we have introduced a decomposition of the identity operator
at each intermediate time $\phi= n\epsilon,\,\, n=1,2,\ldots ,
N-1$; denoted the matrix element $\bra \b\v_n| \exp iH \epsilon|
\b\v_{n-1}\ket$ by $U_{\b\v_n\b\v_{n-1}}$; and, for notational
simplicity, set $\v_f = \b\v_N$ and $\v_i = \b\v_0$. The division
of $\I$ provides a skeletonization of the time interval. An
assignment $\sigma_N = (\b\v_N, \ldots ,\b\v_0)$ of volumes to the
$N+1$ time instants $\phi = \epsilon n$ can be regarded as a
discrete history associated with this skeletonization since one
can envision the system as hopping from $\b\v_{n-1}$ to $\b\v_n$.
The transition amplitude is thus given by a sum of amplitudes over
these discrete histories, a la Feynman:
\be \label{soh1} A(\v_f,\varphi;\,\v_i,0) = \sum_{\sigma_N}\,
A(\sigma_N)\quad\quad {\rm with}\quad A(\sigma_N) =
U_{\b\v_N\b\v_{N-1}}\, U_{\b\v_{N-1}\b\v_{N-2}}\, \ldots
U_{\b\v_2\b\v_1}\,U_{\b\v_1\b\v_0} \, . \ee

\section{Vertex Expansion: Reorganizing the Sum Over Histories}
\label{s3}

In this section we will reorganize the sum (\ref{soh1}) in the
spirit of SFMs and cast it as a `vertex expansion'. First note that
along a path $\sigma_N$, the volume $\b\v$ is allowed to remain
constant along a number of time steps, then jump to another value,
where it could again remain constant for a certain number of time
steps, and so on. The first key idea is to group paths according to
the number of \emph{volume transitions} rather than time steps. Let
us then consider a path $\sigma_N^M$ which involves $M$ volume
transitions (clearly, $M\le N$):
\be \sigma_N^M = (\, \v_M,\ldots ,\v_M;\,\v_{M-1},\ldots,\v_{M-1};\,
\ldots\,\, \ldots ,\v_2;\overbrace{\v_1,\ldots ,\v_1;\,
\underbrace{\v_0,\ldots,\v_0}_{N_1}}^{N_2}\,)\, . \ee
Thus, the volume changes from $\v_{m-1}$ to $\v_{m}$ at time $\p =
N_m\epsilon$ and remains $\v_m$ till time $\p = N_{m+1}\, \epsilon$.
(Note that $\v_m$ is distinct from $\b\v_m$ used in section
\ref{s2}: while $\v_m$ is the volume after the $m$-th \emph{volume
transition} along the given discrete path, $\b\v_m$ is the volume at
the end of the $m$-th \emph{time interval}, i.e., at $\p =
m\epsilon$.)

These discrete histories can be labelled more transparently by two
ordered sequences
\be \sigma_N^M = \{\,(\v_M, \v_{M-1},\ldots ,\v_1, \v_0);\,\, (N_M,
N_{M-1}, \ldots , N_2, N_1)\,\},\,\,\, \v_{m}\not=\v_{m-1}, N_m >
N_{m-1}. \ee
Note that while no two \emph{consecutive} volume values can be
equal, a given volume value can repeat in the sequence; $\v_{m}$ can
equal some $\v_n$ if $n\not=m\pm 1$. The probability amplitude for
such a history $\sigma_N^M$ is given by:
\be A(\sigma_N^M) = [U_{\v_M\v_M}]^{N-N_M-1}\,\, U_{\v_M
\v_{M-1}}\,\, \ldots\, [U_{\v_1\v_1}]^{N_2-N_1-1}\,\,
U_{{\v_1}{\v_0}}\,\, [U_{\v_0\v_0}]^{N_1-1} \ee
The second key idea is to carry out the sum over all these
amplitudes in three steps. First we keep the ordered set of volumes
$(\v_M, \ldots , \v_0)$ fixed, but allow the volume transitions to
occur at \emph{any} value $\p = n\epsilon$ in the interval $\I$,
subject only to the constraint that the $m$-th transition occurs
before the ($m$+1)-th for all $m$. The sum of amplitudes over this
group of histories is given by
\be \label{1} A_N(\v_M, \ldots, \v_0) = \sum_{N_M=M}^{N-1} \,\,
\sum_{N_{M-1}=M-1}^{N_M-1}\, \ldots\, \sum_{N_1=1}^{N_2-1}\,\,
A(\sigma^M_{N}). \ee
Next we sum over all possible intermediate values of $\v_m$ such
that $\v_m\not=\v_{m-1}$, keeping $\v_0=\v_i,\, \v_M=\v_f$ to
obtain the amplitude $A(M)$ associated with the set of all paths
in which there are precisely $M$ volume transitions:
\be \label{2} A_N(M) = \sum_{\substack{\v_{M-1},\ldots,\v_{1} \\
\nu_m \neq \nu_{m+1}}} \; A_N(\v_M, \ldots, \v_0) \ee
Finally the total amplitude $A(\v_f,\p;\,\v_i,0)$ is obtained by
summing over all volume transitions that are permissible within our
initially fixed skeletonization with $N$ time steps:
\be \label{3} A(\v_f,\varphi;\,\v_i,0) = \sum_{M=0}^{N} A_N(M)
\equiv \sum_{M=0}^N\,\, \Big[\sum_{\substack{\v_{M-1},\ldots,\v_{1} \\
\nu_m \neq \nu_{m+1}}} \; A_N(\v_M,\ldots ,\v_0)\, \Big]\, .
 \ee

Recall, however, that the Hamiltonian theory implies
$A(\v_f,\varphi;\,\v_i,0) = \bra\v_f|e^{iH\varphi}|\v_i\ket$.
Hence the value of the amplitude (\ref{3}) does not depend on $N$
at all; the skeletanization was introduced just to express this
well-defined amplitude as a sum over histories. Thus, while the
range of $M$ in the sum and the amplitude $A_N(M)$ in (\ref{3})
both depend on $N$, the sum does not.

The third key idea is to get rid of the skeletonization altogether
by taking the limit as $N$ goes to infinity, to express the total
transition amplitude as a vertex expansion in the spirit of the
timeless framework of spin-foams. To carry out this step, we first
note that a straightforward but non-trivial calculation \cite{ach}
shows that $\lim_{N\to\infty} A_N(\v_M,\ldots, \v_0)$ exists and is
given by:
\ba\label{lim1} A(\v_M,\ldots,\v_0) &:=& \lim_{N\to\infty}\,
A_N(\v_M,\ldots, \v_0)\nonumber\\
&=& \int_0^\varphi \!\dd\p_M\, \int_0^{\p_M} \!\dd\p_{M-1}\, \ldots
\int_0^{\p_2} \!\dd\p_{1}\,\,\, A(\v_M, \ldots ,\v_0; \p_M,\ldots\,
,\p_1)\, ,\ea
where,
\ba \label{lim2} A(\v_M, \ldots ,\v_0;\p_M,\ldots ,\p_1) &:=&
e^{i(\varphi-\p_M)H_{\v_M\v_M}}\,\, (iH_{\v_M\v_{M_1}})\,\,
e^{i(\p_M-\p_{M-1})H_{\v_{M-1}\v_{M-1}}}\,\times \nonumber\\
&& \ldots\,\, e^{i(\p_2-\p_1)H_{\v_1\v_1}}\,\,
(iH_{\v_1\v_{0}})\,\, e^{i\p_1 H_{\v_0\v_0}} \ea
The structure of these equations can be understood as follows. In
the limit $N \rightarrow \infty$, the length $\epsilon = \varphi/N$
of the elementary time intervals goes to zero and discrete sums in
(\ref{1}) are replaced by continuous integrals. Eq (\ref{lim1}) says
that the final, i.e., $M$-th volume transition can occur anywhere in
the interval $\I$, the ($M$-1)-th transition can occur anywhere
before the $M$-th, and so on. In passing from (\ref{1}) to
(\ref{lim2}), factors like $[U_{\v_M\v_M}]^{N_M-N_{M-1}} = [1+ \\
i((N_M- N_{M-1})\epsilon) H_{\v_M\v_M} + \ldots ]$ go over to
$e^{i(\p_M-\p_{M-1})H_{\v_M\v_M}}$ while factors like $U_{\v_M
\v_{M-1}} = i\epsilon H_{\v_M\v_{M-1}} + O(\epsilon^2)$ go over to
$iH_{\v_M\v_{M-1}}$.

It is trivial to carry out the integrations over $\p_m$ in
(\ref{lim1}) and express $A(\v_M, \ldots ,\v_0)$ just in terms of
matrix elements of $H$. For simplicity, let us consider the case
when all of $(\v_M, \ldots ,\v_0)$ are distinct. Then, we have:
\be \label{int1}A(\v_M, \ldots ,\v_0) = H_{\v_M\v_{M-1}}\, \ldots\,
H_{\v_1,\v_0} \,\, \sum_{m=0}^M \f{e^{i\varphi\,H_{\v_m
\v_m}}}{\prod_{\substack{j=0\\ j\not= m}}^M\, (H_{\v_m\v_m} -
H_{\v_j\v_j})}\, .\ee
All matrix elements $H_{\v_m,\v_n}$ can be computed explicitly
\cite{ach}. Finally, since $\lim_{N\to\infty} A_N(\v_M,\ldots,
\v_0)$ exists, in the limit (\ref{3}) becomes simply
\be \label{total} A(\v_f,\varphi;\,\v_i,0) = \sum_{M=0}^{\infty}
A(M)\quad\quad
{\rm where,}\quad A(M) = \sum_{\substack{\v_{M-1},\ldots,\v_{1} \\
\nu_m \neq \nu_{m+1}}} \;  A(\v_M, \ldots ,\v_0) \, .  \ee

Eq (\ref{total}) mimics the vertex expansion of SFMs. More
precisely, the parallels are as follows. The analog of the manifold
$M$ with boundaries $S_i,S_f$ in SFMs is the manifold $\V\times\I$,
where $\V$ is the elementary cell in LQC and $\I$
the closed interval bounded by $\p=0$ and $\p=\varphi$. 
The analog of the dual-triangulation in SFMs is just a `vertical'
line in $\V\times\I$ with $M$ marked points or `vertices' (not
including the two end-points of $\I$). What matters is the number
$M$; the precise location of vertices is irrelevant. Coloring of
the dual-triangulation in SFMs corresponds to an ordered
assignment $(\v_M,\v_{M-1}, \ldots \v_1,\v_0)$ of volumes to edges
bounded by these marked points (subject only to the constraints
$\v_M= \v_f,\,\, \v_0=\v_i$ and $\v_m \not= \v_{m-1}$). Each
vertex signals a change in the physical volume along the quantum
history. The probability amplitude associated with the given
coloring is $A(\v_M, \ldots ,\v_0)$ and a sum over colorings
yields the amplitude $A(M)$ associated with the triangulation with
$M$ `vertices'. The total amplitude $A(\v_f,\varphi;\,\v_i,0)$ is
given by a sum (\ref{total}) over these $M$-vertex amplitudes.

To conclude this section, we emphasize that the result was
\emph{derived} from a Hamiltonian theory. We did not postulate that
the left side of (\ref{total}) is given by a formal path integral.
Rather, a rigorously developed Hamiltonian theory guaranteed that
the left side is well-defined and provided the expression
(\ref{lqc}) for it. We simply recast this expression as a vertex
expansion. 

\section{Vertex Expansion as a Perturbation Series}
\label{s4}

We will now show that the expression (\ref{total}) of the transition
amplitude can also be obtained using a specific perturbative
expansion. Structurally, this derivation of the vertex expansion is
reminiscent of the perturbative strategy used in group field theory
(see, e.g., \cite{gft1,gft2}).

Let us begin by considering the diagonal and off-diagonal parts $D$
and $K$ of the `Hamiltonian' $H$ in the basis $|\v = 4n\lo\ket$,
defined by their matrix elements:
\be \label{DK} D_{\nu' \nu} = H_{\nu \nu}\, \delta_{\nu' \nu} ,
\quad\quad K_{\nu' \nu}  =  \left\{ \begin{array}{ll} H_{\nu' \nu}
& \quad \nu' \neq \nu \\
0 & \quad \nu' = \nu
\end{array}\right.
\ee
Clearly $H= D+K$. The idea is to think of $D$ as the main part of
$H$ and $K$ as a perturbation. To implement it, introduce a
1-parameter family of operators $H_{\l}= D + \lambda K$ as an
intermediate mathematical step. The parameter $\lambda$ will simply
serve as a marker to keep track of powers of $K$ in the perturbative
expansion; we will have to set $\lambda=1$ at the end of the
calculation.

Following the textbook procedure, let us define the `interaction
Hamiltonian' as
\be H_I(\p)=e^{-i D\p}\,\, \lambda K \,\, e^{i D\p}. \ee
Then the evolution in the interaction picture is dictated by the
operator
\be \tilde{U}_\lambda(\p)= e^{-i D\p} e^{iH_{\l}\,\p}\, , \quad\quad
{\rm satisfying} \quad \f{\dd\tilde{U}_\l(\p)}{\dd \p} = i
H_I(\p)\tilde{U}_\lambda(\p)\, , \ee
whose solution is given by a time-ordered exponential:
\ba
\tilde{U}_\lambda(\varphi) & = & \mathcal{T}\,\,
e^{i \int_0^\varphi H_I(\p)\dd\p}\nonumber \\
& = & \sum_{M=0}^{\infty} \int_{0}^{\varphi} \dd\p_{M}
\,\int_{0}^{\p_{M}}\! \dd\p_{M-1}\ldots \int_{0}^{\p_2}\! \dd\p_1
\,\,\,  [iH_I(\p_{M})]\, ...\, [iH_I(\p_{1})]\, .\label{utilde}
\ea

Next, let us express the evolution operator as $U_\lambda(\varphi) =
e^{i D\varphi}\tilde{U}_\lambda(\varphi)$, with
$\tilde{U}_\lambda(\varphi)$ given by (\ref{utilde}), take matrix
element between initial and final states, $|\nu_i \equiv \nu_0\ket$
and $|\nu_f \equiv \nu_M\ket$, and write out explicitly the product
of the $H_I$'s. The result is
\ba A_\lambda(\nu_f,\varphi; \nu_i, 0) = \sum_{M=0}^{\infty}\,
\int_{0}^{\varphi}\! \dd\p_{M}&\,...\,& \int_{0}^{\p_2}\! \dd\p_1
\sum_{\v_{M-1},\,\ldots,\,\v_{1}} [e^{i (\varphi-\p_M)
D_{\nu_M \nu_M}}]\,\,(i\lambda K_{\nu_M\nu_{M-1}})\,\,\times\nonumber\\
&&[e^{i(\p_{M}-\p_{M-1}) D_{\nu_{M-1} \nu_{M-1}}}]\, \ldots\,
(i\lambda K_{\v_1\v_0})\,\, [e^{i\p_1 D_{\v_0\v_0}}]\, . \ea
We can now replace $D$ and $K$ by their definition (\ref{DK}).
Because $K$ has no diagonal matrix elements, only the terms with
$\v_m\not=\v_{m+1}$ contribute to the sum and the sum becomes
\be \label{final1} A_\lambda(\nu_f,\varphi; \nu_i, 0) =
\sum_{M=0}^{\infty}\, \lambda^M \,
\Big[\sum_{\substack{\v_{M-1},\ldots,\v_{1} \\
\nu_m \neq \nu_{m+1}}} \; A(\v_M,\ldots, \v_0)\,\Big]\, , \ee
where $A(\v_M,\ldots,\v_0)$ is defined in (\ref{int1}). If we now
set $\lambda=1$, Eq. (\ref{final1}) reduces to Eq. (\ref{total})
obtained independently in section \ref{s3}.

Thus, by formally regarding the off-diagonal piece of the
Hamiltonian as a perturbation of the diagonal piece we have
obtained an independent derivation of the vertex expansion of the
amplitude $A_\lambda(\nu_f,\varphi; \nu_i, 0)$ as a power series
expansion in $\lambda$, the power of $\lambda$ serving as a
book-keeping device to keep track of the order in the vertex
expansion. In this sense this alternate derivation is analogous to
the vertex expansion obtained using group field theory.

\section{Group Averaging}
\label{s5}

We began our discussion by carrying out a deparametrization (using
$\p$ as the relational time variable) because this is the procedure
used in LQC to extract physics from the quantum theory. Spin foams
on the other hand are based on a timeless framework. Furthermore, a
convenient deparametrization is not always available even in
cosmology if we allow the scalar field to have general potentials or
the gravitational field to admit inhomogeneities. In this case we
have to return to the full constraint and construct the physical
Hilbert space differently. Then the basic object of interest is not
a transition amplitude but the physical inner product. In LQC, it is
given by the well-known group averaging procedure:
\be \bra\v_f,\p_f|\v_i,\p_i\ket_{\rm phys} =
\big[(\v_f,\p_f|\textstyle{\int_{-\infty}^{\infty}}\dd\alpha\,
e^{i\alpha\,C}\big]\,|\v_i,\p_i\ket \ee
where $C= p^2_\p/\hbar^2 - \Theta$ is the constraint operator and
the round bracket on the right side of the equation denotes a
`generalized bra', an element of the algebraic dual (called ${\rm
Cyl}^\star$ in the literature) of a suitable dense subspace of $\Hk$.%
\footnote{There is some freedom in the definition of the action of
elements of ${\rm Cyl}^\star$. In LQC, this freedom is used to
simplify the expression of the physical inner product \cite{aps1}
and the subsequent action of Dirac observables on $\Hp$. We will use
the same conventions here.}
Formally, integration over the `lapse' $\alpha$ introduces the
factor $\delta(C)$ that is necessary to extract physical states from
kinematical ones and also yields the physical inner product between
the resulting physical states. This procedure can be carried out in
detail \cite{aps1}. If we restrict ourselves to the positive part of
the spectrum of $\h{p}_\p$ ---or, to `positive frequency' physical
states--- as in LQC, \emph{the physical inner product is given
precisely by the transition amplitude} $A(\v_f,\p_f; \v_i,\p_i)$ we
focused on in the last three sections (which, in turn, reproduces
(\ref{norm})). On the `negative frequency' solutions it is given by
the complex conjugate, $[A(\v_f,\p_f; \v_i,\p_i)]^\star$ (because
the matrix elements $H_{\v_m,\v_n}$ and $\Theta_{\v_m,\v_n}$ are all
real.) If we allow both, then the inner product is always real:
$\bra\v_f,\p_f|\v_i,\p_i\ket_{\rm phys} = A(\v_f,\p_f; \v_i,\p_i) +
[A(\v_f,\p_f; \v_i,\p_i)]^\star$. Thus, the physical inner product,
the key object in the timeless framework, can be readily constructed
from the transition amplitude, the key object in the deparameterized
framework.

We can also use the procedure followed in section \ref{s3} or
\ref{s4} to carry out group averaging directly, without any
reference to $\p$ as relational time, to express the physical inner
product as a vertex expansion. The two procedures yield the same
result which is of course equivalent to (\ref{total}). However,
somewhat surprisingly, order by order, it is distinct from
(\ref{total}). For definiteness, let us use the perturbative method
of section \ref{s4}. The key object now is the operator $e^{i\alpha
C}$. As before, let us introduce a 1-parameter family of operators
$C(\ul\l)= p_\p^2/\hbar^2 - \Theta(\ul{\l})$ as follows. Set
$\Theta(\ul\l) = \ub{D} +\ul\l\,\ub{K}$ where $\ub{D}$ is the
diagonal part of $\Theta$  and $\ub{K}$ the off-diagonal part. Then,
using the interaction picture one can again expand out the operator
$e^{i\alpha\,C(\ul{\l})}$ and take its matrix elements to obtain the
$\ul{\l}$-physical inner product to obtain
\be\label{final2} \bra \v_f,\p_f|\v_i,\p_i\ket_{{\rm phy},\,\ul{\l}}
= \sum_{M=0}^{\infty}\, \ul{\l}^M \,
\Big[\sum_{\v_{M-1},\ldots,\v_{1}} \ub{A}(\v_M,\ldots,
\v_0)\,\Big]\, , \ee
where
\be \label{int2} \ub{A}(\v_M,\ldots, \v_0) = \Theta_{\v_M\v_{M-1}}\,
\ldots\, \Theta_{\v_1,\v_0} \,\, \sum_{m=0}^M \f{e^{i(\p_f-\p_i)
\sqrt{\Theta_{\v_m \v_m}}}} {\prod_{\substack{j=0\\
j\not= m}}^M\, (\Theta_{\v_m\v_m} - \Theta_{\v_j\v_j})}\, .\ee
(As before, for simplicity we have assumed that $(\v_{M}, \ldots
,\v_0)$ are distinct.) Again, $\ul{\l}$ is only a book-keeping
device for the intermediate perturbative expansion and the physical
inner product is obtained by setting $\ul{\l}=1$ in the final
result.

Note that Eq (\ref{final2}) is identical to Eq (\ref{final1})
(except for underbars) and Eq (\ref{int2}) has the same form as Eq
(\ref{int1}). However, while Eq (\ref{int1}) contains matrix
elements of $H$, Eq (\ref{int2}) contains matrix elements of $\Theta
= H^2$. This fact leads to two important differences. First
$\sqrt{\Theta_{\v_m\v_m}}$, the square root of the matrix element of
$\Theta$, is distinct from $H_{\v_m\v_m}$, the matrix element of the
square root of $\Theta$. Second, because the off diagonal elements
$\Theta_{\v_m\v_n}$ in Eq (\ref{int2}) are non-zero only if
$\v_m=\v_n\pm 4\lo$, consecutive $\v_m$ in second sum in Eq
(\ref{final2}) can differ only by $\pm 4\lo$. There is no such
simplification in Eq (\ref{final1}). Because of these differences,
although the physical inner product obtained by group averaging is
related in a simple manner to the transition amplitude, the vertex
expansion obtained in this section is completely different from that
obtained in the last two sections. If we were to terminate the sum
at any finite order, the results would not be simply related.

\section{Discussion}
\label{s6}

Let us start with a brief summary. In section \ref{s2} we began with
Hamiltonian LQC, divided the time interval into $N$ segments and
expressed the transition amplitude $A(\v_f,\p_f;\, v_i,\p_i)$ as a
sum (\ref{soh1}) over discrete histories. In section \ref{s3} we
reorganized this sum emphasizing volume transitions and took the
$N\to \infty$ limit to get rid of the skeletonization of the time
interval. This led us to the expression (\ref{final1}) of the
transition amplitude. The $M$-th term in this expansion corresponds
to a sum over all histories in which there are precisely $M$ volume
transitions, allowed to occur at \emph{any} time in the interval
$(\p_f,\p_i)$. Therefore the expansion resembles the vertex
expansion of SFMs. In section \ref{s4} we showed that the same
vertex expansion can be arrived at by formally splitting the
Hamiltonian $H$ into a main part $D$ and a `perturbation' $\l K$ and
expanding the transition amplitude using standard perturbation
theory in the interaction picture. (Here the `coupling constant'
$\l$ was introduced just as a mathematical label to keep track of
the number of vertices in various terms and we have to set $\l=1$ at
the end to recover the physical transition amplitude). This
expansion in powers of $\l$ resembles the vertex expansion in group
field theory. Finally in section \ref{s5} we returned to the
expression of the physical inner product that one begins with in the
group averaging procedure. This features the exponential of the
constraint operator $C \equiv p_\p^2 -\Theta$, where $\Theta = H^2$.
One can carry out a perturbative expansion as in section \ref{s4} by
formally writing $\Theta$ into a main part $\ub{D}$ and a
perturbation $\ul{\l} \ub{K}$ to arrive at an expression
(\ref{final2}) of the inner product as a perturbation series in
$\ul{\l}$. The standard group averaging procedure implies that (for
`positive frequency' physical states) the physical inner product in
the timeless framework equals the transition amplitude in the
deparameterized framework used in LQC \cite{aps1}. But whereas the
perturbative expansion (\ref{final1}) involves the matrix elements
of $H$, (\ref{final2}) involves the matrix elements of $\Theta$ (and
their square-roots). Thus, while the sum yields the same quantity,
if we were to truncate the perturbation series to any finite order
one would obtain distinct results. Thus, somewhat surprisingly there
are two \emph{distinct} natural vertex expansions, one descending
from the deparameterized theory and the other from the timeless
framework of group averaging.

While our final result yields vertex expansions in the spirit of
SFMs, we did not begin with a SFM and arrive at the vertex expansion
by a symmetry reduction, e.g., by summing over the degrees of
freedom other than the total volume. Our procedure is much more
modest: As is usual in LQC, we carried out the symmetry reduction at
the \emph{classical} level by partial gauge fixing, constructed the
Hamiltonian quantum theory and used it to obtain vertex expansions.
Also, so far, vertex expansions have been discussed in SFMs only for
source-free gravity, while the presence of a scalar field played a
key role in LQC. Nonetheless our results provide strong support for
the paradigm underlying SFMs. In addition, since we have an exactly
soluble, concrete example, we can use it to analyze the status of
open issues.

First, as hoped in SFMs, the physical inner product can indeed be
expressed as a vertex expansion. The inner product is defined
independently and this well-defined quantity is merely expanded out
as a convenient series. 
Second, the expectation that the expansion should be derivable from
a suitable Hamiltonian theory has been realized. In addition, as one
would expect from \cite{rr}, each vertex can be thought of as
emerging from the action of the Hamiltonian operator. Finally, in
our decomposition $H = D +\l K$ (or $\Theta = \ub{D} +
\ul{\l}\ub{K}$) one can think of $D$ as the free part of the
Hamiltonian because it does not change the volume and $K$ as the
interaction part because it does. Thus, as in group field theory,
the factors of $\l$ are associated with `interaction' piece of the
Hamiltonian (which is responsible for the volume transitions in
LQC).

An issue that is often raised in the literature on SFMs  is whether
the physical inner product is really given by just summing over
triangulations, each with a finite number of vertices, or if one
should take a ``continuum'' limit at the end as in, e.g., lattice
QCD. If one \emph{defines} the inner product as a discretized path
integral, the answer is not a priori obvious. However, since we
began with a well-controlled Hamiltonian theory, in the LQC example,
the answer \emph{is} clear. We did not have to take the limit; the
discrete sum provided the exact answer. It is also instructive to
note that we were led to  two natural vertex expansions. As we
remarked in section \ref{s1}, in SFMs one has to introduce a vertex
amplitude and, while it is constrained by several requirements, we
do not have a statement of uniqueness. In LQC we were able to obtain
two distinct expansions, one using the deparameterized theory and
the other using the timeless framework. Are there perhaps similar
inequivalent vertex expansions in more complicated models ---or even
the full theory--- each tailored to making an aspect of the theory
more transparent? Next, we saw that the inner product between
physical states extracted from the kinematic basis vectors $|\v,
\p\ket$ are in general complex (as is generally true for constrained
systems). However, if we were to enlarge the physical Hilbert space
allowing for both `positive \emph{and} negative frequency'
solutions, they become real. In SFMs the obvious analog of the LQC
`positive frequency' restriction is a choice of an orientation.
Currently, the sum involves both orientations and the inner products
between the physical states extracted from spin network states are
all real. The LQC analysis naturally raises a question: Should one
also impose a suitable restriction allowing, e.g., only those
histories with only one orientation \cite{do}? Or, does correct
quantum physics require us to have only real physical scalar
products in this basis? If so, why is there a qualitative difference
in LQC? Finally, if one regards group field theory as fundamental,
rather than just a convenient computational tool to arrive at the
spin foam vertex expansion, then one is led to take the coupling
constant $\lambda$ as a physical parameter which can run with the
renormalization group flow. However, its interpretation has been
elusive. A detailed examination of the LQC example shows that it is
naturally tied with the cosmological constant. If this were to hold
also in the full theory, one may have a dynamical tool to analyze
why the cosmological constant is so small in the low energy regime.
These and several other issues will be discussed in detail in
\cite{ach}.

\bigskip

\textbf{Acknowledgments:} We would like to thank Jerzy Lewandowski,
Daniele Oriti and Carlo Rovelli for their comments. This work was
supported in part by the NSF grant PHY0854743 and the Eberly
research funds of Penn State.

\end{document}